\documentstyle[osa,manuscript,psfig]{revtex}

\begin{document}

\titlepage
\title{
	\hskip 7.0truecm {{{\normalsize FUB-HEP/95-21 (revised version)}}}\\
	\vspace {1.2truecm}
Probing dissociation of space-like photons in deep-inelastic 
lepton-nucleon scattering} 
\author {C. Boros, Liang Zuo-tang and Meng Ta-chung}
\address {Institut f\"ur theoretische Physik,
  Freie Universit\"at Berlin \\
 Arnimallee 14, 14195 Berlin, Germany}
	
\maketitle

\begin{abstract}
It is shown that the hadronic dissociation
of space-like $(Q^2>0)$ photons can
be directly probed 
by performing measurements 
in the fragmentation region of
transversely polarized and unpolarized proton 
beams at the electron-proton collider HERA. 
Measurements of momentum-distributions in the photon-fragmentation 
region in normal and in LRG (large rapidity gap) events are also 
suggested --- especially when the scattered proton or neutron in the 
proton-beam direction is tagged. It is pointed out that 
such distributions can yield useful information 
on the mechanisms of hadronic fragmentation in general, and 
answer the following questions in particular: 
Is  the well-known hypothesis of limiting fragmentation (HFL) valid in 
color-exchange, in flavor-exchange, or only in 
vacuum-quantum-number-exchange processes ?

\end{abstract}

\newpage

It is known already for a long time that 
hadronic dissociation of space-like
photons may play a
significant role in deep-inelastic 
lepton-hadron scattering --- especially
in diffractive processes [1,2]. 
People seem to agree [1-18] that, 
viewed from the hadron- or nucleus-target, 
not only real, but also space-like photons 
$(Q^2\equiv -q^2>0$, 
where $q$ is the four-momentum of such a photon) 
may exhibit hadronic structure. 
But, as far as the following  questions are concerned, 
different theoretical models [1-18] 
seem to give different answers. 
How do such hadronic dissociation processes 
depend on the standard kinematic
variables of deep-inelastic 
lepton-nucleon scattering, namely on $Q^2$ and
on $x_B\equiv -q^2/(2pq)$,  
where $p$ is the four-momentum of 
the struck nucleon? 
In particular, how do such virtual photons 
behave in the small $x_B$ and large $Q^2$
regions where 
the photons are far from their mass-shells, for example
$x_B=10^{-2}$ and $Q^2\approx 2000$ (GeV/c)$^2$ 
--- values which can be readily reached at HERA?
We think it would be useful to answer these questions directly --- 
by performing measurements in deep-inelastic lepton-hadron 
scattering processes.  
The reasons are the following:

I. Viewed from the rest frame of the struck nucleon
mentioned above, the lifetime $\tau_\gamma$ of the virtual hadronic
system (of quark-antiquark pair or pairs) 
is of the order $2\nu/Q^2 =1/(Mx_B$), where $\nu$ is the
photon-energy and $M$ is the proton-mass. 
This means, the corresponding longitudinal  dimension --- 
also known as the formation/coherence length ---
of such a virtual state is 
of the order 100 Fermis for $x_B=10^{-2}$.
Furthermore, we note that $\tau_\gamma$ is  a function of $x_B$,  
independent of $Q^2$. 
Does it imply that 
the hadronic dissociation 
of a photon always takes place --- independent of 
its virtuality $Q^2$?
Is it true that, in high-energy deep-inelastic lepton-hadron 
scattering processes, we are practically  
{\it always} dealing with hadron-hadron collisions, 
when $x_B$ is sufficiently small ($x_B\sim 10^{-2}$, say)? 

II. Some of the dynamical models 
based on such a photon-dissociation picture
(See e.g. Refs.12-18 and the papers cited there) 
have been used to describe quantitatively 
the proton structure function
$F_2^p(x_B,Q^2)$ in the small $x_B$ region 
and the obtained results are in reasonable agreement 
with the existing data [19,20]. 
Can we, on the basis of this agreement, say: 
``Experiments show that space-like photons $\gamma^*(Q^2)$ 
always dissociate into hadronic systems 
--- independent of their virtualities ($Q^2$-values)'' ?

III. Until now, besides the measurements 
of the proton structure function 
$F_2^p (x_B,Q^2)$ in the small-$x_B$ region, 
the only ``direct experimental tests" 
for such pictures and/or scenarios
have been deep-inelastic lepton-nucleus 
scattering experiments (See e.g. Ref. 18), 
in which information on photon-dissociation 
can be extracted from the reactions between the
space-like photons and (light and heavy) nuclei. 
Having seen (See e.g. Ref.18 and the references given there)
how many different
theories reproduce the well-known 
lepton-nucleus-collision data, and how many
different versions of the ''generalized 
vector-meson model"  fit 
the existing data for shadowing and/or
anti-shadowing effects, 
it seems rather natural to ask: 
Wouldn't it be useful also to have 
methods alternative 
to lepton-nucleus collision for this purpose?

IV. It is known that, in the usual parton description of $F_2^p(x_B,Q^2)$, 
the question whether photons dissociate has been bypassed by describing the 
reaction mechanism in a fast-moving frame 
--- the appropriately chosen ``infinite momentum frame'' 
--- in which the virtual photon carries little energy and thus 
the lifetime for the virtual 
hadronic state in the dissociation is short. 
But, as has already been pointed out more then twenty years 
ago by Nieh[1] the following is also true: 
Because of the shortened time scale for interaction in
that frame, the dissociated hadronic state may not need to live very
long to become effectively important; 
and hence, it seems there does not 
exist strong theoretical reason for 
making the assumption that the photon-dissociation mechanism 
is not important in the ``infinite momentum'' frame. 
Can we say:`` The question whether photon dissociation 
takes place is independent of the  reference frame in which the 
observation is made''?


In this paper we propose to check hadronic dissociation
of space-like photons by performing
inclusive measurements in 
the fragmentation region of transversely polarized
and unpolarized proton beams at HERA  
and by comparing the results with those obtained in 
corresponding hadron-hadron collisions. 
This is because we think, if we know what the characteristic 
features of hadron-hadron collisions are, we may check 
whether/when such typical features  occur in the same manner in 
$\gamma^*(Q^2)$-hadron collision processes. 
By doing so, we can find out experimentally whether/when $\gamma^*(Q^2)$ 
behaves like a hadron. In other words, we 
can say whether/when hadronic dissociation of $\gamma^*(Q^2)$ 
takes place. Hence, in this connection, it seems useful to know the 
following: 
Are there indeed such characteristic features for 
hadron-hadron collisions ? 
Is there experimental evidence that $\gamma^*(Q^2)$ for 
some $Q^2$-values indeed behaves like a hadron ?

(a). A number of high-energy 
hadron-hadron collision experiments [21-25] --- 
especially proton-proton
collision experiments at CERN-ISR [21-24] 
show  that the particles observed in the 
fragmentation regions play an extremely important role in 
understanding their production mechanisms. 
First of all, it is observed [21] that  
{\it limiting (i.e. energy-independent), rapidity-distributions exist in 
the rest frame of the fragmenting beam.} 
It is  observed [22,23] in particular that,  
while $\pi^+$, $\pi^-$ 
and $K^+$-mesons with not too small transverse
momenta $(p_\bot \ge 0.5~GeV/c$, say) 
significantly contribute to 
the projectile fragmentation
region (Feynman-x $x_F\ge 0.4)$, 
{\it the $K^-$-mesons do not.}  
Furthermore, it is observed [24-27] 
that hyperons, in particular $\Lambda^0$'s, 
produced in fixed target hadron-hadron and
hadron-nucleus collisions are {\it polarized}, 
although neither 
the projectile nor the target is polarized. 
This polarization is {\it independent of the 
incident energy} and 
{\it it exists only in the fragmentation region 
of the projectile-proton!}
The question whether such polarization phenomena also exists 
in lepton-lepton-collisions has also been discussed a long time 
ago (see Ref.28 and the references given therein). It is known 
in particular 
that the possible existence 
of $\Lambda$-polarization
has been carefully searched in  
electron-positron collisions, and no evidence 
 has been found [28].

(b). High-energy hadron-hadron collision experiments 
in which {\it the projectile-hadron
is polarized transversely to 
the scattering plane} have been performed [29].
{\it Significant left-right-asymmetries (up to 40\% ) 
with the following properties 
have been observed [29] in inclusive 
$\pi^+,\pi^-,\pi^0,\eta^0$ and $\Lambda^0$
production processes} : 
First, {\it these asymmetries are $x_F$-dependent}: 
They are different from 
zero in the fragmentation region,
and {\it only} in the fragmentation region
$(x_F\ge 0.4)$ of the transversely polarized projectile-hadron.
Second, {\it they are  flavor-dependent}:  
The asymmetries for $\pi^+,\pi^-$ and $\pi^0$ are very much
different from one another.  
Third, {\it they strongly depend on 
the quantum numbers of the projectile}: 
The observed asymmetries for proton and 
those for anti-proton are very much different from
one another. 
Existence of such single-spin asymmetry effects 
in lepton-lepton 
or lepton-hadron collision processes is not known. 

(c). High-energy photo- and electro-production 
experiments [18]
 with fixed proton-target and
nuclear targets show that 
{\it real $(Q^2=0)$ and 
nearly real ($Q^2\le$ 1 or 2 GeV$^2$/c$^2$, say) 
photons behave like vector mesons 
such as $\rho^0,\omega,\phi,J/\psi$ etc} [4,2,10,11,17,18].

From the results mentioned in (a) and (b) we see 
that the projectile-proton   
in high-energy hadron-hadron and hadron-nucleus collisions 
exhibits striking 
features {\it in its fragmentation region}.
These features 
(e.g. the $\Lambda$-polarization) 
observed {\it in the projectile fragmentation region} 
depends only on the quantum-numbers of the projectile, 
{\it but not on that of the hadronic target} 
(e.g. different nuclei). 
From the result mentioned in (c), we see that 
real ($Q^2=0$) and space-like ($Q^2>0$) photons 
indeed behave like hadrons for small 
$Q^2$-values. 
Having these experimental facts in mind, 
it seems natural to ask: 
Can we use the phenomena which have been observed ---- and only observed
--- in hadron-hadron collisions as characteristic features for
hadron-hadron collision processes ? 
Shall we  see  
{\it such characteristic features in 
proton's fragmentation region} in high-energy 
proton-$\gamma ^*(Q^2)$ collisions 
if $\gamma^*(Q^2)$ indeed behaves like a hadron? 
Can we use the observation of 
{\it such characteristic features in proton's 
fragmentation region} 
in proton-$\gamma^*(Q^2)$ collisions 
as signal for hadronic dissociation of $\gamma^*(Q^2)$ for given values 
of $Q^2$ ? 
It seems reasonable and useful to adopt the following standpoint: 
The phenomena which have been observed in hadron-hadron collisions and 
observed only 
in such collisions can be, and should be, considered as 
characteristic properties of hadron-hadron collision processes. In this 
sense, we propose to use the proton in $\gamma^*(Q^2)$-proton collisions 
as a ``sensor'' 
--- as an ``instrument'' --- 
to find out the following: 
Does the  space-like photon $\gamma^*(Q^2)$ with given 
virtuality $Q^2$ act as a hadron ?  
It should be mentioned that, although
the experimental facts listed in (a) and (b) can be 
understood in terms of a relativistic model 
(See Appendix and the references given there); but what we 
wish to find out here is merely whether/when 
$\gamma^*(Q^2)$-proton scattering show the same characteristic features 
as those observed in hadron-proton scattering.  

To be more precise, 
we propose to perform  single-particle 
inclusive measurements in the fragmentation region
of the transversely polarized proton $p(\uparrow )$ 
and/or in  that of the unpolarized protons $p$ at DESY-HERA
in the small-$x_B$ region for different $Q^2$-values, and 
to compare the obtained
results with those obtained 
in the corresponding hadron-hadron collisions.
We note: What we suggest to measure and  to compare  are {\it not} 
the absolute values of the cross sections but rather the 
$\Lambda$-polarization $P_{\Lambda}$ or 
the left-right asymmetry $A_N$   
 in $\gamma^*(Q^2)+p$  processes. 
The quantities $P_\Lambda$ and $A_N$ are ratios of 
the difference and the sum of 
such cross sections. Hence, in these quantities, 
the $1/Q^2$ factors due to the transverse 
geometrical size of $\gamma^*(Q^2)$ are completely cancelled out.

In order to demonstrate in a quantitative manner   
how the $Q^2$-dependence of such dissociation 
processes may manifest itself, we 
examine the $F_2^p(x_B,Q^2)$-data [19,20] in the 
small-$x_B$ region. 
In Fig.1, we separate the well-known vector-dominance 
contribution (See e.g. 17,18 and the references cited there) 
from ``the rest'' which may be identified as 
``the part due to quark-antiquark continuum'' 
or ``the rest of the contributions due 
to the generalized vector-dominance model'', 
and we consider the following two extreme possibilities which correspond to
two very much different physical pictures:  
(i) The hadronic dissociation of virtual space-like
$(Q^2>0)$ photons take place 
for all possible $Q^2$-values. 
In particular, ``the rest'' mentioned above 
is independent of $Q^2$. In other words, in this picture $\gamma^*(Q^2)$ should 
always be considered as a hadronic system --- independent of $Q^2$. 
(ii) The hadronic dissociation
of such photons depends very much on $Q^2$. 
In terms of a two-component picture (See e.g.Ref.8 )
 the virtual photon $\gamma^*(Q^2)$ 
is considered to be either in the
``bare photon'' state or 
in a hadronically dissociated state (``hadronic cloud'') described 
by the vector-dominance 
model [4,2,10,11,17,18 and the papers cited therein]. 
In other words, in this picture ``the rest'' mentioned above is 
strongly $Q^2$-dependent.

Let us first look at the 
left-right asymmetry data [29] for 
$\pi^{\pm}$-production in 
$p(\uparrow )+p$ and see what we may obtain by 
replacing the unpolarized proton-target $p$ by 
a photon with given $Q^2$, $\gamma ^*(Q^2)$. 
It is clear that the corresponding asymmetry which we denote by 
$A_N(x_F,Q^2)$ will have the following properties:  
If scenario (i) is correct, we shall see no 
change in $A_N(x_F,Q^2)$ by varying $Q^2$. 
If scenario (ii) is true, there will be a 
significant $Q^2$-dependence. 
This is shown in Fig.2. 
Similar effects are expected also for $K^+$-mesons.

In this connection, 
it should also be mentioned that such studies can be
performed, even when the detector
is not able to differentiate between
pions and kaons and/or to identify $\Lambda^0$'s. 
Due to the experimental fact that  
the left-right asymmetries $A_N$ of $\pi^+$ and $\pi^-$ produced
in the fragmentation region of 
transversely polarized protons have different
signs, and the fact that the produced mesons 
are predominantly pions, 
(we recall that the forward protons can be identified by the leading
proton spectrometer LPS at HERA), 
{\it a significant left-right asymmetry 
in electric charge is expected in
the inclusive production processes 
$p(\uparrow) +\gamma^*(Q^2) \to$ charged
mesons $+X$}, provided that 
the virtual photons $\gamma^*(Q^2)$ dissociate
hadronically. 

We next consider the $\Lambda$-polarization 
$P_\Lambda(x_F,Q^2)$ 
in the process $p+\gamma^*(Q^2)\to \Lambda +X$ 
in which unpolarized proton beam is used. 
Also here, we expect to see {\it no $Q^2$-dependence 
for scenario (i) but a significant $Q^2$-dependence 
for scenario (ii). }
This is shown in Fig.3.

What do we expect to see in the fragmentation region of 
$\gamma^*(Q^2)$ in the above-mentioned proton-$\gamma^*(Q^2)$ 
collision processes, when $\gamma^*(Q^2)$ dissociates hadronically ? In
this connection, it is useful to recall that one 
of the most 
striking features of high-energy hadron-hadron collisions is 
the existence of limiting distributions for hadron-hadron-collisions as 
have been predicted in the late 1960`s by Benecke, Chou, Yang and Yen
[30]. 
Hence, if $\gamma^*(Q^2)$ indeed behaves like a hadron, we expect to see
that the momentum-distributions of the
$\gamma^*(Q^2)$-fragmentation-products exhibit limiting behavior at 
sufficiently high energies. This can for example be done by varying 
the proton beam energy (820 GeV and 410 GeV, say) for fixed values 
of $x_B$ and $Q^2$. 
Furthermore, 
 we think it would be useful to perform the following measurements: 
First, identify the forward proton 
(e.g. with the leading proton spectrometer LPS at HERA), 
make sure that all the produced hadron 
have large-rapidity gaps with respect to this proton, 
and measure in the fragmentation region of 
$\gamma^*(Q^2)$ the momentum-distributions of the 
hadrons at fixed  $x_B$ - and  
$Q^2$-values in order to see whether 
limiting distributions indeed exist. 
Next, identify the forwards going neutron 
(e.g. with  the forward neutron calorimeter FNC at HERA) which 
is well-separated by 
large rapidity gaps with respect to the produced hadrons, 
and measure the momentum-distributions of the $\gamma^*(Q^2)$-fragments 
 to check if the hypothesis of limiting fragmentation HLF [30,31] is valid. 
After these have been done, 
compare the results with those obtained in 
lepton-proton scattering events without distinct large rapidity gaps 
(i.e. the ``normal'' events). 
The reasons for such measurements are not difficult to guess: 
Having in mind that
single diffractive hadron-hadron
scattering is nothing else but 
a special case of fragmentation processes
for which HLF is valid [30], it is clear that  
such measurements and comparisons can/should be 
useful in clarifying the following questions:  
Do we see limiting fragmentation of  space-like 
photons $\gamma^*(Q^2)$ in the entire kinematical range of 
$x_B$ and $Q^2$, or only in the small $x_B$ and low $Q^2$ region ? Is 
 HLF valid only when 
``vacuum quantum numbers'' are exchanged between the two colliding
objects;  
or is it also valid when flavor(s) or color(s) are exchanged ? 

\bigskip

The experiments proposed in this paper have been presented and discussed
on various occasions in Berlin and Hamburg. The authors thank 
the organizers and participants of the seminars and 
workshop-sessions, especially M. Derrick, D.H.E. Gro\ss{}, H. Haessler, 
H. Jung, S. Nurushev and R. Rittel for helpful discussions. 
This work was supported in part by 
Deutsche Forschungsgemeinschaft (DFG: Me 470/7-1).

\vskip 1.0truecm

\noindent
{\bf {\large APPENDIX}}

The purpose of this appendix is to point out 
--- from a theoretical point of view [30,31,34-40] --- 
that the quoted[21-27] and the proposed experiments 
are closely related to one another, 
and that the proposed ``sensor'' 
is expected to work well. 

We recall that, according to our present knowledge, 
baryons ($p, \Lambda$ etc.) 
are made out of three valence quarks, 
vector- and scalar-mesons ($\rho, \omega, \phi, ..., \pi^\pm, K^\pm$
etc.) consist of a valence quark and a valence antiquark 
--- together with in general a large number 
of sea quark and antiseaquark pairs. 
Universal  distributions of all these 
quarks/antiquarks have been extracted from 
deep-inelastic lepton-hadron scattering 
and lepton-pair production experiments. 
It is a remarkable fact that the 
CERN-ISR experiments[22,23,24] mentioned 
in (a) in the text show the following:
The $x_F$-distributions for 
$\pi^+(\equiv u\bar d),\pi^-(\equiv d\bar u)$ and
$K^+(\equiv u\bar s)$ are very much 
the same as the distributions for the valence-quarks
$u(x_F),d(x_F)$ and $u(x_F)$ respectively. 
The $x_F$-distribution for 
$K^-(\equiv \bar d s)$, however, behaves very much different 
from those of $\pi^+,\pi^-$ and $K^+$.
In fact, it falls off extremely fast 
in the projectile fragmentation region $x_F\ge 0.4$; 
and its magnitude is  
only about 1/20 of that
for $K^+$ at $x_F\approx 0.6$ and 1/100 
of that for $K^+$ at $x_F\approx 0.7$. 
In other words, $K^-(\equiv \bar d s)$ 
is almost absent in the fragmentation region of the
proton $p(\equiv uud)$. 
Taken together with the  empirical
fact[21] concerning limiting fragmentation [30]
these properties clear show that 
{\it the valence quarks of the projectile hadron play a dominating
role in meson-production in the projectile fragmentation region!}.
To be more precise, 
it can be, and has been, explicitly shown [34,35] that, 
while the $x_F$-distribution for $\pi ^+$, $\pi^-$ and 
$K^+$ are the convolutions of the $u_v$, $d_v$ and 
$u_v$ valence quarks with those of the corresponding 
antiseaquarks $\bar d_s$, $\bar u_s$ and $\bar s_s$, 
the $x_F$-distribution for $K^-$ is that of the 
a sea quark $s_s$ and an antiseaquark $\bar d_s$. 

Next, we examine the data for the produced $\pi ^+$ 
and $\pi ^-$ in the single-spin reactions  
$p(\uparrow )+p\to \pi^\pm+X$ mentioned in (b). 
It has been shown[36-38] that the observed 
left-right-asymmetry can be readily described in the 
framework of a relativistic quark-model in which 
the observed $\pi ^+$ and $\pi ^-$ are respectively 
the fusion-products of the valence quarks 
$u_v$ and $d_v$ of $p(\uparrow )$ and 
antiseaquarks $\bar d_s$ and $\bar u_s$. 
Here, the geometrical properties of the hadrons   
in particular  the  
surface-effects play an important role. 
Furthermore, it has been shown[39] that the left-right 
asymmetry of $\Lambda$'s in 
$p(\uparrow )+p\to \Lambda  +X$ as well as the polarization in 
$p+p\to \Lambda +X$ and 
$p+\mbox {nucleus} \to \Lambda  +X$ 
mentioned in (a) can also be understood [40] in terms 
of the above-mentioned relativistic quark model. 

The CERN-ISR proton-proton collision 
experiment performed by Bellitini et al [21] explicitly 
show that the hypothesis of 
limiting fragmentation [30]
hadron-hadron collisions is valid. 
This hypothesis is based on a geometrical picture (See Ref.30 and the
papers cited there), 
in which the colliding hadrons at sufficiently high energies 
simply ``go through each other''. 
During such collision processes, the colliding hadrons 
in general become excited, 
and subsequently decay independently from each other. 
Single diffractive scattering, in which one of the 
colliding hadron remains unchanged, 
is nothing else but a special case. 
What do we know about the reaction mechanism(s) of 
limiting fragmentation of hadrons ? We know that, in such processes, 
 only a rather limited amount of energy-momentum
transfer takes place. But, the question whether /which 
 other physical quantities (quantum numbers) can be or should be 
 exchanged is yet unanswered. 
By performing the measurements   
proposed in this paper, 
we expect to see the following more clearly: 
(i) the relationship between hadron-hadron 
and virtual photon-hadron scattering; 
(ii) the relationship between 
large rapidity gap  events[32,33] 
and normal events in lepton-proton scattering; 
and (iii) the mechanisms 
of color-, flavor- and vacuum quantum-number exchange in general, 
and the relationship between such quantum number-exchange and 
the validity of HLF [30,31] in particular.

\newpage

\noindent
{\large Figures}

\vskip 0.5truecm

\noindent
Fig.1. Structure function $F_2^p(x_B,Q^2)$ as a function of $Q^2$. 
The data-points are taken from [19,20]; and they are parametrized 
(shown as solid line) in order to carry out the quantitative calculation 
mentioned in the text. 
The dashed line is the contribution from the 
vector meson dominance. The difference, which is called ``the rest'', is
shown as dotted line. 

\vskip 0.5truecm

\noindent
Fig.2.
Left-right asymmetry for pion-production in 
$p(\uparrow ) +\gamma^*  \rightarrow \pi^\pm +X$ as a function of 
$x_F$ at different values of $Q^2$. The data are for $p(\uparrow ) +p
\rightarrow \pi^\pm +X$ and are taken from Ref. [29]. 
See text for more details.

\vskip 0.5truecm

\noindent
Fig.3. 
Polarization for $\Lambda$-production in 
$p +\gamma^*  \rightarrow \Lambda  +X$ as a function of 
$x_F$ at different $Q^2$. The data are 
taken from Ref. [24-27]. 
See text for more details.

\vskip 0.5truecm

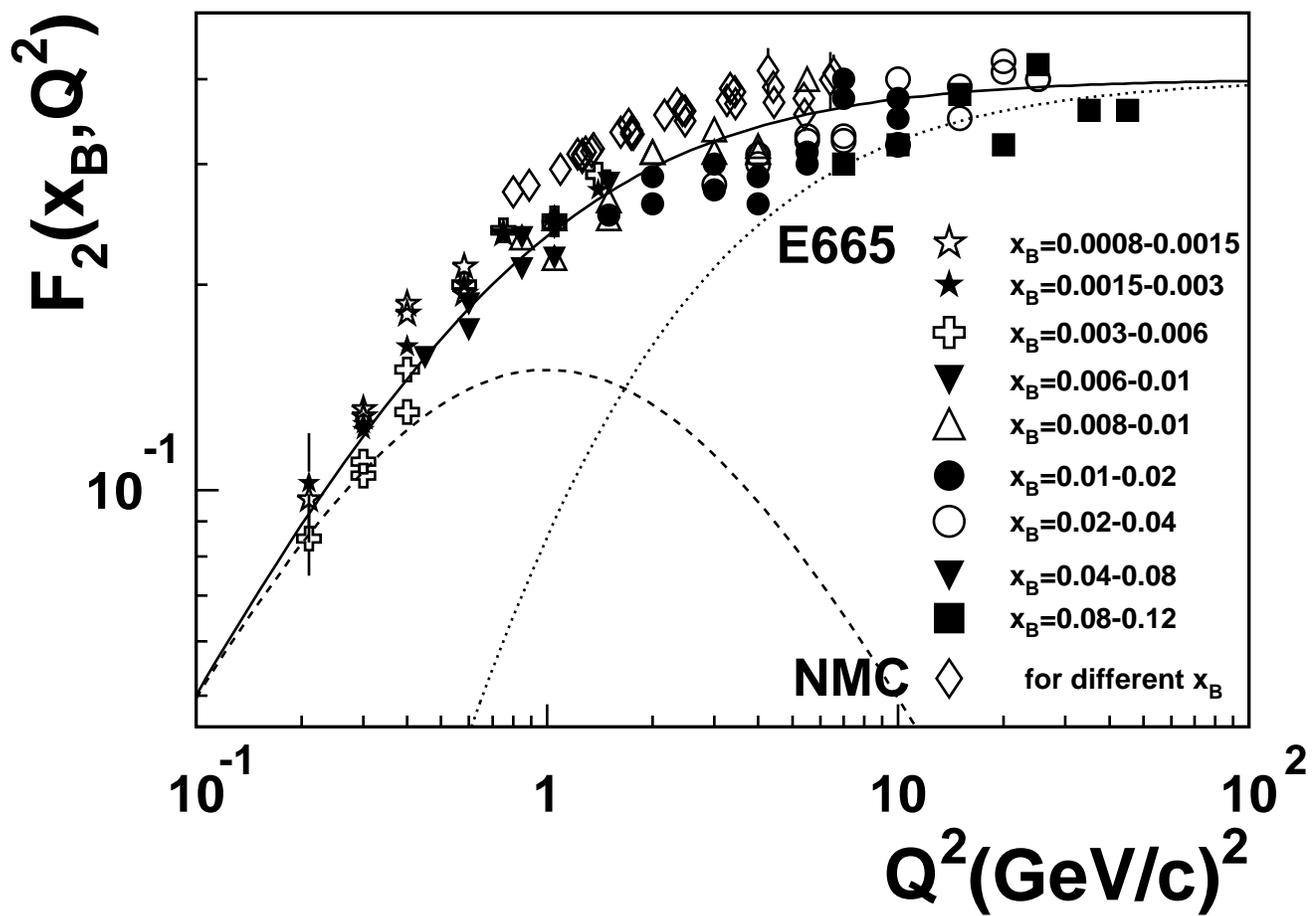
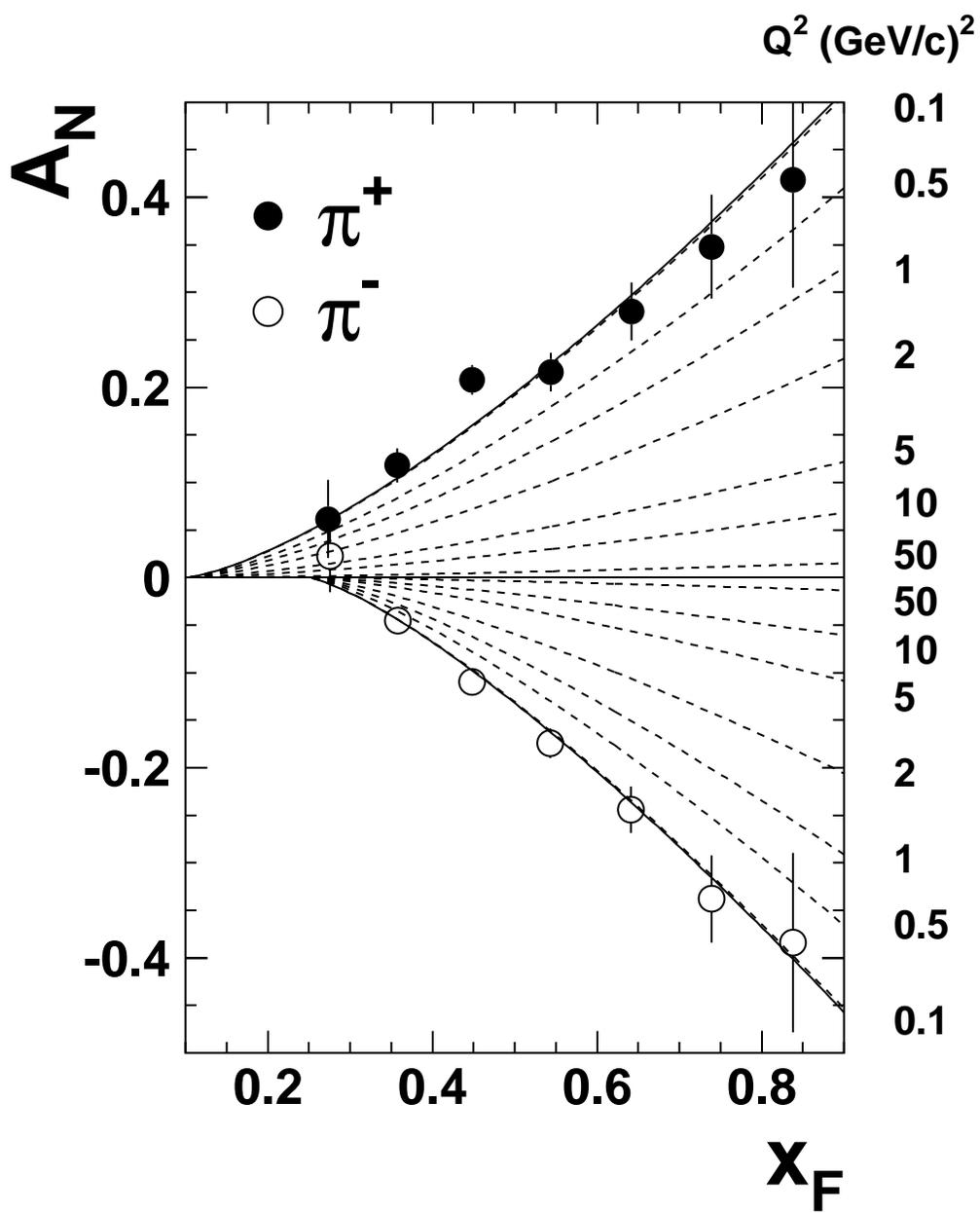
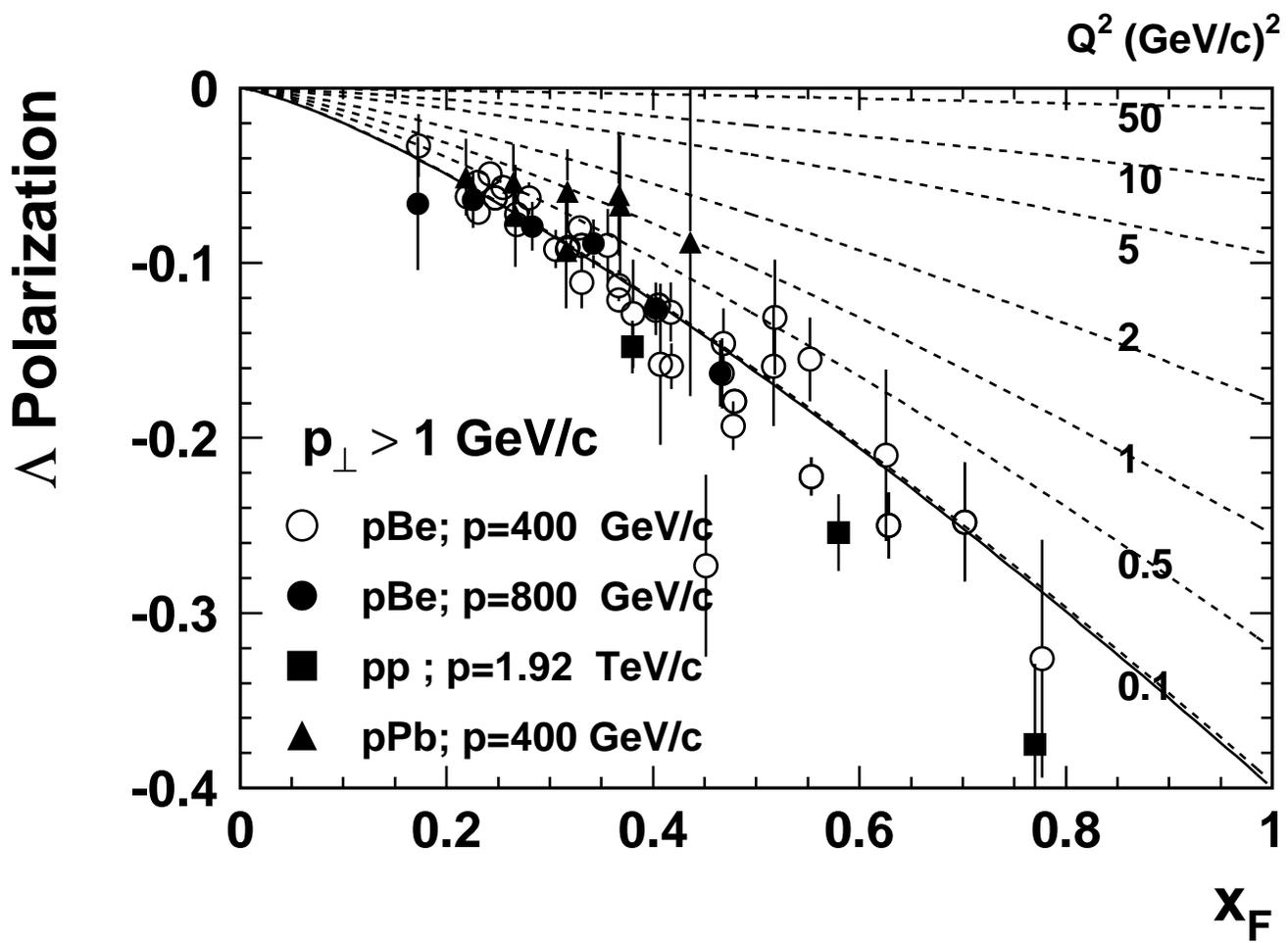
\end{document}